\documentclass[twocolumn,pre,showpacs,superscriptaddress,floatfix]{revtex4-1}
\usepackage{graphicx}
\usepackage{amsmath}
\usepackage{amssymb}
\usepackage{color}
\usepackage{bm}

\begin{document}
\title{Degree-ordered percolation on hierarchical scale-free network}

\author{Hyun Keun Lee}
\affiliation{Department of Physics and Astronomy, Seoul National University,
Seoul 151-747, Korea}
\affiliation{School of Physics, Korea Institute for Advanced Study, Seoul
130-722, Korea}
\author{Pyoung-Seop Shim}
\affiliation{Department of Physics, University of Seoul, Seoul 130-743,
Korea}
\author{Jae Dong Noh}
\affiliation{Department of Physics, University of Seoul, Seoul 130-743,
Korea}
\affiliation{School of Physics, Korea Institute for Advanced Study,
Seoul 130-722, Korea}
\date{\today}

\begin{abstract}
We investigate the critical phenomena of the degree-ordered
percolation~(DOP)
model on the hierarchical $(u,v)$ flower network. Using the
renormalization-group like procedure, we derive the recursion relations for
the percolating probability and the percolation order parameter, from which
the percolation threshold and the critical exponents are obtained.
When $u\neq 1$, the DOP critical behavior turns out to be identical to that
of the bond percolation with a shifted nonzero percolation threshold.
When $u=1$, the DOP and the bond percolation have the same vanishing
percolation threshold but the critical behaviors are different.
Implication to an epidemic spreading phenomenon is discussed.
\end{abstract}


\maketitle

\section{introduction}
Percolation refers to the emergence of a giant cluster
during a process where nodes or links are being added.
Since a global connectivity is one of the key factors for a proper function 
of a network, the percolation transition has been received much attention 
in various contexts. This raises the question regarding the nature of
the percolation transition in different applications.
The nature depends on whether nodes or links
are added in a random or strategic 
way~\cite{Cohen00,Callaway00,Albert00,Gallos05,Hooyberghs10},
whether underlying networks are correlated or 
uncorrelated~\cite{Noh07,SWKim08,Goltsev08}, and whether they are 
static or growing~\cite{Callaway01,JKim02}.

One of the applications where the percolation plays an important role
is the susceptible-infected-removed~(SIR) model on complex
networks~\cite{Newman02}.
The SIR model describes an
epidemic spreading among individuals which are in a susceptible,
infected, or removed state. An infected individual infects a
neighboring susceptible one or recovers acquiring immunity.
Suppose that one individual is infected while the others are in the
susceptible state initially. Then, after a transient period, the system ends
up with a state consisting of susceptible and removed individuals. The
cluster of removed individuals thus obtained 
is equivalent to a bond percolation
cluster~\cite{Newman02}. Hence the epidemic transition is characterized by
the percolation transition.

In this paper, we investigate the percolation threshold and
the critical behavior of a degree-ordered
percolation~(DOP) in which nodes are occupied in the descending order
of their degree. Higher degree nodes are occupied first.
The DOP was introduced in Ref.~\cite{LeePRE2013} in order to resolve a
controversy concerning the epidemic threshold
for an infection rate $\lambda$
of the susceptible-infected-susceptible~(SIS) model.
In contrast to the SIR model, infected individuals become susceptible
again after recovering. Thus, the SIS model does not correspond to an
ordinary percolation model.

The SIS model was studied in the framework of the heterogeneous
mean-field~(HMF) theory in which node
connectivity was treated in an annealed way~\cite{Pastor-Satorras01}.
It predicts a threshold $\lambda_c^{\rm HMF}$
as a function of cumulants of the degree,
which may be zero or nonzero depending on the degree distribution. Later on,
a more refined quenched mean-field~(QMF) theory was proposed, in which node
connectivity was treated in a quenched way~\cite{nep,nep2,nep3}.
The QMF yields that the density $\rho$ of infected
nodes becomes nonzero at $\lambda_c^{\rm QMF}
\lesssim 1/\sqrt{k_{\rm max}}$ where
$k_{\rm max}$ is the highest degree among all nodes. It leads researchers to
conclude that the epidemic threshold vanishes in any networks where the maximum
degree diverges in the infinite-size limit~\cite{nep2,nep3}.

This conclusion might be true if $\rho$ would remain finite 
in the infinite-size limit
for $\lambda>\lambda_c^{\rm QMF}$.
However, it was found that the infection of the
QMF theory can be localized around the hub with the maximum
degree for a finite $\lambda$ larger than
$\lambda_c^{\rm QMF}$~\cite{GoltsevPRL2012}.
The localization property was more elaborated
in Ref.~\cite{LeePRE2013}. As $\lambda$ increases, more localized infections
appear around nodes whose degree $k$ satisfies $k \gtrsim 1/\lambda^2$.
Localized infections are finite-sized, hence decay after a transient period.
The true epidemic transition will take place when the localized
infections percolate at $\lambda_c>\lambda_c^{\rm QMF}$.
Note that localized infections appear successively around nodes
in the degree-descending order. Thus, the DOP can provide a valuable
information on the epidemic threshold. It was claimed that
a nonzero DOP threshold implies a nonzero epidemic
threshold~\cite{LeePRE2013},
which was confirmed numerically in a model network.
The controversy is still under debate~\cite{Boguna13,LeeCmmt2013}.

A complex network is vulnerable to a targeted attack to higher degree
nodes~\cite{Albert00}. The DOP deals with the opposite situation where 
higher degree nodes are preferred. 
It is an interesting question whether the preference
can make the percolation threshold vanish in complex networks.
Furthermore, considering the relevance to the SIS model,
it is important to establish firmly
the percolation threshold and the critical property of the DOP in complex
networks whose maximum degree diverges in the infinite size limit.
For this reason, we investigate the DOP in the $(u,v)$ flower
networks~\cite{uvf}. The $(u,v)$ flowers have a hierarchical structure,
which allows an analytic approach. It is best suited for our purpose.

This paper is organized as follows:
In Sec.~\ref{sec:threshold}, we apply the DOP to the hierarchical
$(u,v)$ flowers and find a recursion relation for the percolating
probability. It allows us to find the percolation threshold and the
finite-size scaling exponent exactly. In Sec.~\ref{sec:op}, we derive a set
of recursion relations for the percolation order parameters, and obtain the
order parameter exponent. We summarize the paper with discussion in
Sec.~\ref{sec:summary}.

\section{Percolation threshold}\label{sec:threshold}

\begin{figure}
\includegraphics*[width=\columnwidth]{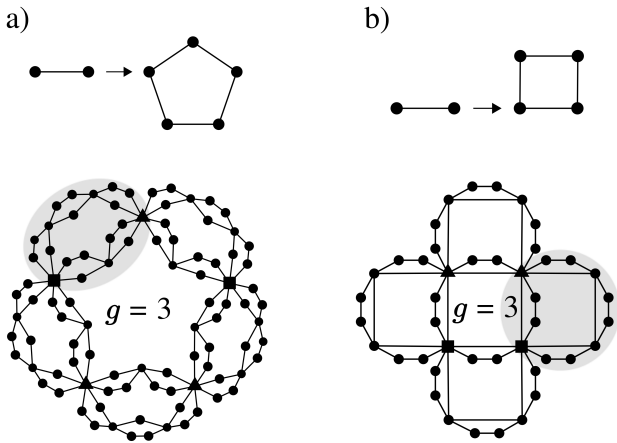}
\caption{Iteration rules and and configurations of $G_{3}$ 
for the $(2,3)$ flower in (a) and $(1,3)$ flower in (b).
The roots and the hubs are denoted by square and triangular symbols, 
respectively. The
shaded area represents one of the descendents of $G_3$.
}\label{fig1}
\end{figure}

The $(u,v)$ flower is a model for a scale-free network with hierarchical
structure~\cite{uvf}. It is generated iteratively starting from a zeroth
generation $G_0$ consisting of two nodes and a link
connecting them. Given a $g$th generation $G_g$, $G_{g+1}$ is
obtained by replacing each link with two linear chains, one of which consists
of $u$ links and $(u-1)$ nodes and the other of which consists of
$v$ links and $(v-1)$ nodes~[see Fig.~\ref{fig1}].
Among others, the initial nodes present in $G_0$ and $G_1$
play an important role
in characterizing the percolation. The two nodes of $G_0$ will be referred
to as roots, and $w\equiv u+v$ nodes of $G_1$ as hubs. A root is also
a hub.

Alternatively, $G_{g+1}$ can be understood as the $w$
copies of $G_g$ that are arranged along a ring with the roots of
adjacent $G_g$s being joined. 
Hereafter, $G_{g}$ embedded in $G_{g+1}$ is called a
descendent, and embedding $G_{g+1}$ is called an ascendent.
The roots of adjacent descendents are merged to become a hub of the ascendent.
Without loss of generality, it is assumed that $u\leq v$.

Summarized below are the structural properties of the $(u,v)$ flower of generation
$g$~\cite{uvf}: (i) The total number of nodes is given by
$N_g = \frac{w-2}{w-1}w^g + \frac{w}{w-1}$, which is obtained from
the recursion relation $N_{g}=w(N_{g-1}-1)$ with $N_0 =2$.
(ii) The degree of nodes takes on a value among
$\{2^1,\cdots,2^g\}$, and the number of nodes having degree
$k=2^l~(l=1,\cdots,g)$
is given by $n_l = (w-2) w^{g-l}$ for $l<g$ and
$n_g=w$. (iii) The cumulative degree distribution is defined as
$P_{\geq}(k) \equiv \sum_{2^l\geq k} n_l/N_g$. 
It decays as $P_{\geq}(k)
\sim k^{1-\gamma}$ with $\gamma=1+\ln w / \ln 2$
in the large $k$ limit.
(iv) The $(u,v)$ flower is a small-world network with the diameter
scaling as $D_g \sim \ln N_g$ for $u=1$ while it is a fractal network
with $D_g \sim N_g^{\ln u / \ln w}$ for $u>1$.

We describe the DOP on the $(u,v)$ flower. Initially all the
nodes are unoccupied or empty. Then, one selects a node with the highest
degree among empty nodes and marks it as an occupied node. In case there are
multiple
candidates, one of them is selected randomly. This procedure is repeated
until the fraction of occupied nodes reaches a prescribed value $p$.

We are interested in the percolating probability denoted by $\Pi=\Pi_g(p)$.
It is defined as the probability that the two roots of $G_g$
are connected via occupied nodes.
One needs to treat the cases with $u=1$ and $u>1$
separately. When $u=1$, the roots are connected via a single link in all
generations.
Thus, the system is percolating if the two roots are occupied.
This implies that $\Pi=1$ for $p\geq n_g/N_g \sim w^{-g}$, so
the percolation threshold vanishes in the infinite-size limit.

When $u>1$, any path connecting the two roots includes
nodes with the minimum degree $k=2$~[see Fig.~\ref{fig1}].
Consequently, all nodes with $k>2$ should be occupied necessarily for
percolation. Remaining nodes have the same degree $k=2$, so they are 
occupied randomly.
Hence, the DOP in the
$(u,v)$ flowers with $u>1$ is equivalent to the ordinary random
node percolation starting from the correlated initial condition
in which all nodes with $k>2$ are occupied.
Let $r$ be the occupation probability of nodes with $k=2$.
It is related to the overall fraction $p$ of occupied nodes through
\begin{equation}\label{p2rg}
p = p_0 + \frac{n_1}{N_g}r
\end{equation}
with $p_0 = 1-n_1/N_g$.
In the infinite-size limit~($g\to\infty$),
Eq.~(\ref{p2rg}) becomes
\begin{equation}\label{p2r}
p = \frac{1+(w-1)r}{w} \ .
\end{equation}
The percolating probability is regarded as a function of $p$ or $r$.

The hierarchical structure allows us to derive a
recursion relation for the percolating probability.
Any path connecting the two roots of $G_g$ necessarily passes through
intermediate $(u-1)$ hubs in one direction or $(v-1)$ hubs in the
other direction. Note that the hubs of an ascendent~($G_g$) 
is the roots of the descendents~($G_{g-1}$). 
Hence, $G_{g}$ is percolated  only when $u$ descendents or 
$v$ descendents are percolated simultaneously.
The roots of all descendents are always occupied with $k>2$ 
and the descendents are independent and statistically identical. 
This yields that
\begin{equation}\label{rr_Pi}
\Pi_g(r) = f(\Pi_{g-1}(r))
\end{equation}
with
\begin{equation}
f(x) \equiv 1 - (1-x^u)(1-x^v) \
\end{equation}
with $g > 1$. 
One need to be cautious in applying the recursion relation to the case with
$g=2$. In the recursion relation, $\Pi_1(r)$ should be interpreted as the
percolating probability of a descendent of $G_2$. The roots of a descendent
is occupied since their degree is $k=4$. Thus, we obtain
\begin{equation}\label{rr_Pi1}
\Pi_1(r) = h(r)
\end{equation}
with
\begin{equation}
h(x) \equiv 1-(1-x^{u-1})(1-x^{v-1}) \ .
\end{equation}
Note that the actual percolating probability of an isolated $G_1$, not a
descendent of $G_2$, is given by $r^2 h(r)$.

\begin{figure}
\includegraphics*[width=\columnwidth]{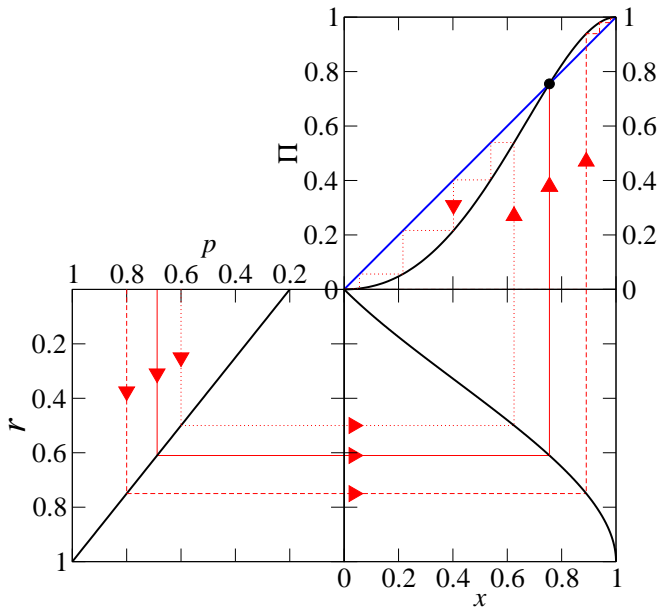}
\caption{(Color online) 
Graphical representation of the recursion relation for $\Pi_g$ at
$(u,v)=(2,3)$.}
\label{fig_iteration}
\end{figure}

\begin{figure}
\includegraphics*[width=\columnwidth]{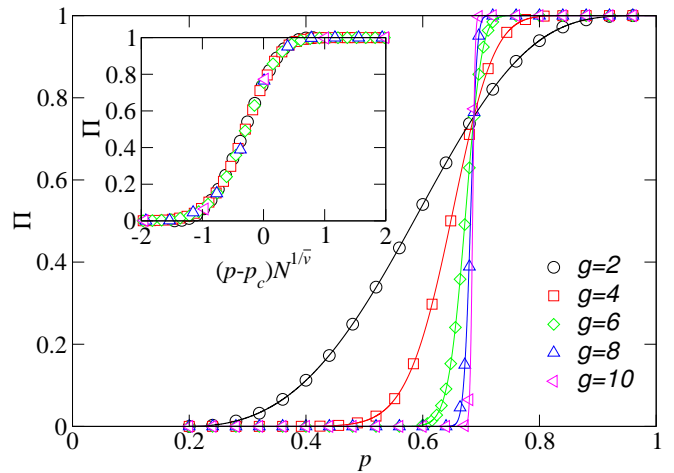}
\caption{(Color online) 
$\Pi_g(p)$ from the recursion relation (solid lines) and from
Monte Carlo simulations~(symbols) in $(2,3)$ flowers.
Inset shows the finite-size-scaling analysis of the percolating
probability.}
\label{Pi_compare}
\end{figure}

The recursion relation is represented graphically in
Fig.~\ref{fig_iteration}. To a given value of $p$,
one obtains $r$ from Eq.~(\ref{p2r}) and $\Pi_1(r)$ from Eq.~(\ref{rr_Pi1}).
Inserting this value into Eq.~(\ref{rr_Pi}) iteratively,
one obtains the percolating probability $\Pi_g$ for all $g>1$.
Figure~\ref{Pi_compare} shows the percolating probability obtained from
the recursion relation in {$(2,3)$} flowers. Also shown are
Monte Carlo simulation data for comparison.

The percolating probability converges to 0~(1) in the $g\to\infty$ limit
when $r$ is less~(greater) than a threshold value $r_c$.
It takes a fixed point value $\Pi_c=x_c$ at the threshold
$r=r_c$ or $p=p_c = (1+(w-1)r_c)/w$.
The threshold is given by
\begin{equation}
r_c = h^{-1}(x_c) \ ,
\end{equation}
where $x_c\neq 0,1$ is the solution of
\begin{equation}
x_c = f(x_c) \ .
\end{equation}
The thresholds at several values of $(u,v)$ are listed in Table~\ref{cridopuv}.

\begin{table}
\caption{Percolation thresholds and critical exponents for the DOP on
$(u,v)$ flowers.
}\label{cridopuv}
\begin{tabular}{ccccccc}
\hline \hline
$~~(u,v)~~$ & $~~~~~\gamma~~~~~$ & $~~~~~p_c~~~~~$ & $~~~~~\Pi_c~~~~~$
& $~~~~~\bar\nu~~~~~$ & $~~~~~\beta~~~~$
\\
\hline
$(2,2)$ 	&	 $3.000$ 	&	 $0.536$ 	&	 $0.618$ 	&	 $3.271$ 	 &	 $0.165$\\
$(2,3)$ 	&	 $3.322$ 	&	 $0.688$ 	&	 $0.755$ 	&	 $3.444$ 	 &	 $0.112$\\
$(3,3)$ 	&	 $3.585$ 	&	 $0.818$ 	&	 $0.848$ 	&	 $3.448$ 	 &	 $0.053$\\
$(2,4)$ 	&	 $3.585$ 	&	 $0.762$ 	&	 $0.819$ 	&	 $3.696$ 	 &	 $0.103$\\
$(3,4)$ 	&	 $3.807$ 	&	 $0.869$ 	&	 $0.890$ 	&	 $3.595$ 	 &	 $0.040$\\
$(4,4)$ 	&	 $4.000$ 	&	 $0.909$ 	&	 $0.921$ 	&	 $3.682$ 	 &	 $0.025$\\
$(2,5)$ 	&	 $3.807$ 	&	 $0.808$ 	&	 $0.857$ 	&	 $3.936$ 	 &	 $0.100$\\
$(3,5)$ 	&	 $4.000$ 	&	 $0.898$ 	&	 $0.913$ 	&	 $3.759$ 	 &	 $0.036$\\
$(4,5)$ 	&	 $4.170$ 	&	 $0.929$ 	&	 $0.938$ 	&	 $3.804$ 	 &	 $0.020$\\
$(5,5)$ 	&	 $4.322$ 	&	 $0.946$ 	&	 $0.951$ 	&	 $3.896$ 	 &	 $0.015$\\
$(2,6)$ 	&	 $4.000$ 	&	 $0.838$ 	&	 $0.881$ 	&	 $4.156$ 	 &	 $0.100$\\
$(3,6)$ 	&	 $4.170$ 	&	 $0.915$ 	&	 $0.929$ 	&	 $3.919$ 	 &	 $0.034$\\
$(4,6)$ 	&	 $4.322$ 	&	 $0.942$ 	&	 $0.949$ 	&	 $3.931$ 	 &	 $0.018$\\
$(5,6)$ 	&	 $4.459$ 	&	 $0.956$ 	&	 $0.960$ 	&	 $3.999$ 	 &	 $0.012$\\
$(6,6)$ 	&	 $4.585$ 	&	 $0.964$ 	&	 $0.967$ 	&	 $4.084$ 	 &	 $0.010$\\

\hline \hline 
\end{tabular}
\end{table}

The percolating probability satisfies the finite-size-scaling~(FSS)
form~\cite{Stauffer94,CompCrit}
\begin{equation}\label{Pi_FSS}
\Pi_g(r) = \mathcal{F}(\epsilon N_g^{1/\bar\nu})
\end{equation}
where $\epsilon = p-p_c$, $\bar\nu$ is the FSS exponent, and
$\mathcal{F}(x)$ is the scaling function. It has the limiting behaviors
$\mathcal{F}(x\to -\infty)\to 0$, $\mathcal{F}(0)=x_c$, and
$\mathcal{F}(x\to\infty)\to 1$. Combining Eqs.~(\ref{rr_Pi}) and
(\ref{Pi_FSS}), we find
\begin{equation}\label{nu}
\bar{\nu} = \frac{\ln w}{\ln f'(x_c)} \ .
\end{equation}
Numerical values of $\bar\nu$ are listed in
Table~\ref{cridopuv}. The FSS form is tested for the $(2,3)$ flowers
in the inset of Fig.~\ref{Pi_compare}.

It is noteworthy that the recursion relation for the percolating probability
of the DOP in the $(u,v)$ flowers with $u\neq 1$ is similar to that
of the ordinary bond percolation
studied in Ref.~\cite{RozenfeldPRE2007}. The difference lies in the fact
that the recursion relation is written in terms of $r$ instead of $p$ and
that $\Pi_1$ has a different form.
The difference only shifts the percolation threshold. The critical
behaviors belong to the same universality class. When $u=1$, the DOP and the
ordinary bond percolation have the same percolation threshold at $p_c=0$.
However, they display different critical phenomena as will be shown
in the following section.

\section{Percolation order parameter}\label{sec:op}

In this section, we investigate the critical scaling of the percolation
order parameter. It is defined as the mean density of nodes
that are connected to any of the hubs.

We first consider the case with $u>1$. Nodes with $k>2$ are
all occupied and remaining nodes with $k=2$ are occupied with probability $r$.
The order parameter $P_\infty=P_\infty(g,r)$ is written as
\begin{equation}\label{P_A}
P_\infty(g,r)= \frac{1}{N_g}
\sum_{m=1}^w
A_{g,m}(r) \ ,
\end{equation}
where $A_{g,m}(r)$ denotes the mean number of nodes
in $G_g$
that are connected to exactly
$m$
hubs
under the condition that the two roots of $G_g$ are occupied.

It is useful to introduce $S=S_g(r)$, the
mean number of nodes in $G_g$ that are connected to only a single root.
It is related to $A_{g,m}$ through the relation
\begin{equation}\label{SfromA}
S_g(r) = \sum_{m=1}^{w} \sigma_{m} A_{g,m}(r) \ ,
\end{equation}
where $\sigma_m$ is the probability that one may find only a single root among
randomly selected $m$ adjacent hubs.
Note there are $w$ different ways in selecting $m$ adjacent hubs except when $m=w$.
Enumerating all the possible cases, we obtain that
\begin{equation}\label{sigma}
\sigma_m = \left\{
   \begin{aligned} &\frac{2m}{w}     &  & (1 \leq m \leq u),  \\
                   &\frac{2u}{w}     &  & (u < m \leq v),  \\
                   &\frac{2(w-m)}{w} &  & (v < m \le w).
   \end{aligned} \right.
\end{equation}
We also introduce $T=T_g(r)$, the mean number of
nodes in $G_g$ that are connected to both roots. It is given by
\begin{equation}\label{TfromA}
T_g(r) = \sum_{m=1}^{w} \tau_m A_{g,m}(r) \ ,
\end{equation}
where $\tau_m$ is the probability that one may find both roots among randomly
selected $m$ adjacent hubs. It is given by
\begin{equation}\label{tau}
\tau_m = \left\{
   \begin{aligned}&  0               &  & (1 \le m \leq u),  \\
                  &\frac{m-u}{w}    &  & (u < m \leq v),  \\
                  &\frac{(2m-w)}{w} & & (v < m \leq w).
   \end{aligned} \right.
\end{equation}

Note that $G_{g+1}$~(an ascendent) consists of $w$ copies of
$G_g$~(descendents) with their roots being joined. Hence, one finds that
$A_{g+1,m} = w (X_{g,m} - c_{g,m})$, where $X_{g,m}$ denotes the
mean number of nodes in one descendent that are connected
to exactly $m$ hubs of the ascendent and $c_{g,m}$ denotes the probability
that a hub of the ascendent is connected to $(m-1)$ other hubs.
Because each hub of the ascendent is shared by two adjacent descendents,
one has to subtract $wc_{g,m}$ from $wX_{g,m}$ in order to compensate for
a double counting.

For $m=1$, we obtain that $X_{g,1} = S_g \overline\Pi_g$,
where $\overline{\Pi}_g \equiv 1-\Pi_g$ with
the percolating probability $\Pi_g$.
The factor $S_g$ accounts for the number of nodes in a descendent
that are connected to one of the roots~(or, equivalently, one of
the hubs of the ascendent), and the factor $\overline\Pi_g$ accounts
for the probability that such nodes are not connected to any other hubs of
the ascendent.
A hub of $G_{g+1}$ is not connected to the adjacent hubs with the
probability $\overline\Pi_g^2$, which yields $c_{g,1}= \overline\Pi_g^2$.
For general $m$, one can easily find that
\begin{equation}\label{AfromST}
A_{g+1,m} = w \left( a_{g,m} S_g + b_{g,m} T_g - c_{g,m} \right)
\end{equation}
where
\begin{equation}
a_{g,m} = \left\{ \begin{aligned}
      & \Pi_g^{m-1} \overline\Pi_g &\quad (1\leq m < w) \\
      & \Pi_g^{w-1} &\quad (m=w)
   \end{aligned}\right. \ ,
\end{equation}
\begin{equation}
b_{g,m} = \left\{ \begin{aligned}
   & (m-1) \Pi_g^{m-2}\overline\Pi_g^2 & (1\leq m<w) \\
   & (w-1) \Pi_g^{w-2}\overline\Pi_g + \Pi_g^{w-1}
   & (m = w)
   \end{aligned}\right. \ ,
\end{equation}
and
\begin{equation}
c_{g,m} = \left\{ \begin{aligned}
   & m \Pi_g^{m-1}\overline\Pi_g^2 & \ (1\leq m < w) \\
   & w \Pi_g^{w-1}\overline\Pi_g +
   \Pi_g^w
   & \ ( m = w)
   \end{aligned} \right. \ .
\end{equation}
Once $\{A_{1,m}\}$ are known, $\{A_{g>1,m}\}$ are obtained by using
Eqs.~(\ref{SfromA}), (\ref{TfromA}), and (\ref{AfromST}).
The order parameter is then evaluated by using Eq.~(\ref{P_A}).

It is more convenient to express the order parameter in terms of $S_g$ and 
$T_g$.
By using Eq.~(\ref{AfromST}) and
$\sum_m a_{g,m} = \sum_m b_{g,m} = \sum_m c_{g,m} = 1$, we find that
\begin{equation}\label{PfromST}
P_{g+1} = \frac{1}{N_{g+1}}\sum_{m=1}^g A_{g+1,m} = \frac{w}{N_{g+1}}
(S_g+T_g-1) \ .
\end{equation}
Combining Eqs.~(\ref{SfromA}), (\ref{TfromA}), and (\ref{AfromST}), one can
derive the following recursion relation
\begin{equation}\label{PtoP}
\left( \begin{array}{cc} S_{g+1} \\ T_{g+1} \end{array} \right) =
w \mathsf{M}_g \left( \begin{array}{cc} S_{g} \\ T_{g} \end{array} \right) -
w \left( \begin{array}{cc}
\sum_m \sigma_m c_{g,m} \\ \sum_m \tau_m c_{g,m}
\end{array} \right)  \ ,
\end{equation}
where $\mathsf{M}_g$ is the $2\times2$ matrix given by
\begin{equation}\label{M_def}
\mathsf{M}_g = \left( \begin{array}{cc}
(\sum_m \sigma_m a_{g,m}) & (\sum_m \sigma_m b_{g,m}) \\
(\sum_m \tau_m a_{g,m})   & (\sum_m \tau_m b_{g,m})
\end{array} \right) \ .
\end{equation}

\begin{figure}
\includegraphics*[width=\columnwidth]{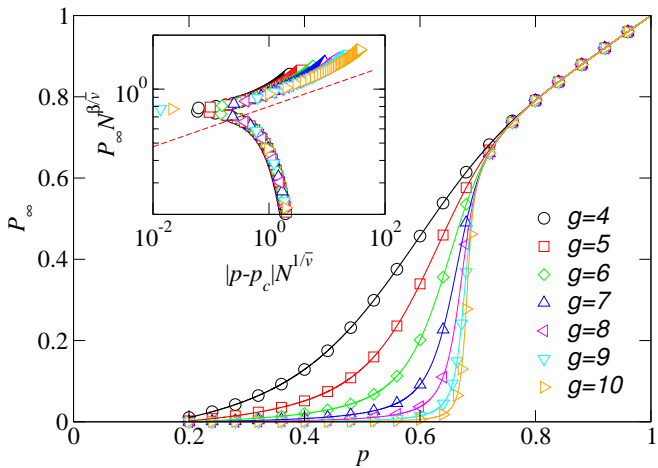}
\caption{(Color online)
The percolation order parameter $P_\infty$ obtained from the recursion
relation~(solid lines) and Monte Carlo simulations~(symbols) in $(2,3)$
flowers. The inset shows the FSS analysis of the order parameter.
The slope of the dashed line is $\beta$.}
\label{fig:P_rec_mc}
\end{figure}

The recursion relation is tested numerically for $(2,3)$ flowers.
In $G_1$ as a descendent of $G_2$, two roots are occupied with probability 1
and other hubs are occupied with probability $r$. It is straightforward to
obtain
that $S_1 = 2(1+2r)(1-r)^2$ and $T_1 = r(3+6r-4r^2)$.
Inserting these into Eqs.~(\ref{PfromST}) and (\ref{PtoP}) iteratively,
we obtain the order parameter in all $g$.
The numerical results are presented and compared with Monte Carlo simulation
results in Fig.~\ref{fig:P_rec_mc}. Both agree with each other perfectly.

The critical scaling behavior of the order parameter is also obtained by
analyzing the recursion relation. The order parameter is
expected to follow the FSS form~\cite{Stauffer94}
\begin{equation}
\label{betaFSS}
P_\infty(p) = N_g^{-\beta/\bar\nu}{\cal G}(\epsilon N^{1/\bar\nu}),
\end{equation}
where $\epsilon = p-p_c$ and the scaling function has the limiting behaviors
$\mathcal{G}(x\to\infty) \sim x^\beta$ and $\mathcal{G}(|x|\ll 1)\sim
\mbox{const.}$ with the order parameter exponent $\beta$. In the preceding
section, we have obtained the critical point $p_c$ or $r_c$ and the FSS
exponent $\bar\nu$. Hence, the exponent $\beta$ can be obtained from the FSS
behavior of the order parameter at $p=p_c$:
\begin{equation}\label{op_c}
P_\infty(p_c) \sim N_g^{-\beta/\bar\nu} \ .
\end{equation}
In the subcritical phase, the number of nodes
connected to the hubs are order of unity with $P_\infty(p<p_c) \sim
N_g^{-1}$. Hence, it is expected that $\beta/\bar\nu<1$.

At the critical point with $\beta/\bar\nu<1$, we can ignore the last terms
in Eqs.~(\ref{PfromST}) and (\ref{PtoP}) and use the
fixed point value $\Pi_c$ instead of $\Pi_g$. Then, the
order parameter scales as
\begin{equation}\label{Pg_res}
P_\infty \sim (w\Lambda_c)^g/N_g \sim \Lambda_c^g
    \sim N_g^{\ln \Lambda_c / \ln w}\ ,
\end{equation}
where $\Lambda_c$ is the
largest eigenvalue of $\mathsf{M}_c = \left. \mathsf{M}_g
\right|_{\Pi_g=\Pi_c}$.
Comparing Eqs.~(\ref{op_c}) and (\ref{Pg_res}), we find that
\begin{equation}
\frac{\beta}{\bar\nu} = -\frac{\ln \Lambda_c}{\ln w} \ .
\end{equation}
The knowledge of $\bar\nu$ and $\beta$ completes the critical behavior of
the DOP on the $(u,v)$ flowers. 

Consider the $(2,3)$ flower as an example. 
We have $\Pi_c \simeq 0.755$ and $\bar\nu \simeq
3.44$ in Table~\ref{cridopuv}. The matrix $\mathsf{M}_c$ is given by
\begin{equation}
\label{m33}
\mathsf{M}_c =
\left(
\begin{array}{ccccc}
\frac{2}{5} & \frac{4}{ 5} & \frac{4}{5} & \frac{2}{5} & 0
\vspace{2mm}
\\
0 & 0 & \frac{1}{5} & \frac{3}{5} & 1
\end{array}
\right)
\left(
\begin{array}{cc}
\bar \Pi_c & 0\vspace{1mm}\\
\Pi_c \bar \Pi_c & \bar \Pi^2_c\vspace{1mm}\\
\Pi^2_c \bar \Pi_c & 2 \Pi_c \bar \Pi^2_c\vspace{1mm}\\
\Pi^3_c \bar \Pi_c & 3 \Pi^2_c \bar \Pi^2_c\vspace{1mm}\\
\Pi^4_c & 4 \Pi^3_c \bar \Pi_c + \Pi^4_c
\end{array}
\right),
\end{equation}
with $\bar\Pi_c = 1-\Pi_c$. It has the largest eigenvalue
$\Lambda\simeq 0.948820 \cdots $, which yields that $\beta/\bar\nu \simeq 0.0326426 \cdots $ and
$\beta \simeq 0.112424 \cdots $.
The inset of Fig.~\ref{fig:P_rec_mc} shows that the numerical data indeed
follow the FSS scaling form.
The critical exponents at several values
of $u$ and $v$ are listed in Table~\ref{cridopuv}.

We add a remark on our recursion relation given in Eqs.~(\ref{PfromST}) and
(\ref{PtoP}) in comparison with that obtained
for the random bond percolation in Ref.~\cite{RozenfeldPRE2007}.
The recursion relations in both studies look
similar to each other except for the terms involving $c_{g,m}$ in
Eq.~(\ref{PtoP}). Those terms account for the double counting of hubs.
Such terms are overlooked in Ref.~\cite{RozenfeldPRE2007}.
Fortunately, they contribute to the order parameter as a subleading
correction term. Hence, the results for the critical exponents in
Ref.~\cite{RozenfeldPRE2007} are valid. The double
counting problem is also noticed in Ref.~\cite{Hasegawa13}.

\begin{figure}
\includegraphics*[width=\columnwidth]{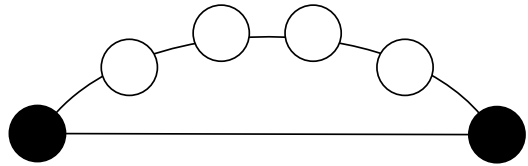}
\caption{A diagram illustrating the unit of $(v-1)$ nodes with
$k=2^l$~(denoted by open symbols) connecting two nodes with $k=2^{l'}$
with $l'>l$~(denoted by closed symbols). }
\label{fig:u1v5_unit}
\end{figure}

Finally, we investigate the nature of the percolation transition
in $(u=1,v)$ flowers, where the percolation transition takes place
at $p=p_c=0$. We describe how the occupied cluster changes its shape as $p$
increases.
When $p=P_{\rm \ge}(k=2^g)$,
all the hubs with $k=2^g$ are occupied and
connected to each other.
Suppose that $P_{\rm \ge}(2^g) < p \leq P_{\rm \ge} (2^{g-1})$ or
$p= P_{\rm \ge}(2^g) + s n_{g-1}/N_g $ with $0<s\leq 1$.
Then, one needs to consider the nodes with $k=2^{g-1}$ additionally.
These nodes form linear chains of length $v-1$ connecting two neighboring
hubs.
Figure~\ref{fig:u1v5_unit} illustrates a unit consisting of a pair of
hubs~(closed symbols) and the linear chain of $(v-1)$
nodes with $k=2^{g-1}$~(open symbols). Those nodes that are occupied and
connected to the hubs contribute to the percolation order parameter. When
$s=1$, all nodes are connected to each other to form a single cluster.

As $p$ increases further, we have a nested structure of the units.
Suppose, in general, that $P_{\rm \ge}(2^{l})<p\leq P_{\rm \ge}(2^{l-1})$ or
$p=P_{\rm \ge}(2^l) + s n_{l-1}/N_g$ with a certain $l$ and $0<s\leq 1$.
All the nodes with $k\geq 2^l$ are occupied with probability 1 and
belong to a single cluster. In addition, nodes with $k=2^{l-1}$ are attached
to every neighboring pair of nodes of higher degree
to form the unit structure in Fig.~\ref{fig:u1v5_unit}~(see also
Fig.~\ref{fig1}).
Given the occupation probability $s$ of the nodes with
$k=2^{g-l}$~(open symbols),
the probability $q_m$ that $m$ nodes among $v-1$ nodes in the unit
are connected to the higher degree nodes~(closed symbols) is given by
\begin{equation}
q_m = \left\{ \begin{aligned}
   & (m+1) s^m (1-s)^2 &\ (m<v-2) \\
   & (m+1) s^m (1-s)   &\ (m=v-2) \\
   & s^m               &\ (m=v-1)
\end{aligned} \right. \ .
\end{equation}
The mean value is given by
\begin{equation}\label{q(s)}
q(s) = \sum_m m q_m = \frac{ (v-3)s^v - (v-1) s^{v-1} + 2s}{1-s} \ .
\end{equation}
Therefore, the percolation order parameter is given by
\begin{equation}\label{P_res_u1}
P_\infty(p) = P_{\rm \ge} (k=2^{l}) + \frac{q(s)}{(v-1)} \frac{n_{l-1}}{N_g} \ ,
\end{equation}
where $l$ is the smallest integer satisfying $p>P_{\rm \ge}(2^l)$ and
$s=(p-P_{\rm \ge}(2^l))/P(2^{l-1})$.

\begin{figure}
\includegraphics*[width=\columnwidth]{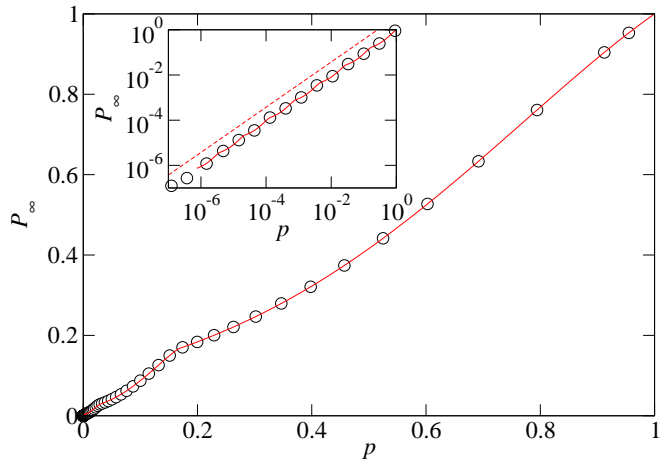}
\caption {(Color online) Order parameter obtained from Monte Carlo
simulations~(symbol) and from Eq.~(\ref{P_res_u1})~(line) in the $(1,5)$
flower in generation $g=9$. The inset shows the plot in the log-log scale.
The straight line of slope 1 is a guide to an eye.}
\label{fig:u5v1}
\end{figure}

When $v=2$ or $v=3$, all nodes selected in the degree descending order are
connected to each other. Indeed, Eq.~(\ref{q(s)}) yields $q(s) = (v-1)s$
and the percolation order parameter becomes $P_\infty(p) = p$.
When $v>3$, $q(s)$ deviates from $(v-1)s$.
Consequently, $P_\infty(p)$ displays an oscillatory behavior with the period
$\Delta p/p \simeq 1/w$ superimposed over the overall behavior $P_\infty(p)=p$.
The percolation order parameter $P_\infty(p)$ calculated from Monte Carlo
simulations is compared with the analytic result in Fig.~\ref{fig:u5v1}.
Both are in perfect agreement with each other.
Therefore, we conclude that the DOP on $(u=1,v)$ flowers has a percolation
threshold at $p_c=0$ and the order parameter exponent is given by
$\beta = 1$.

\section{Summary and discussions}\label{sec:summary}
We have introduced the DOP and investigated the critical phenomena in
the $(u,v)$ flower networks. The hierarchical structure of the $(u,v)$
flowers allows us to find the percolation threshold and the critical
exponents exactly. When $u=1$, the percolation transition takes place at
$p=p_c=0$ with the order parameter exponent $\beta=1$. When $u\neq 1$, the
percolation threshold is nonzero and the critical exponents take the
nontrivial values depending on $u$ and $v$. The percolation thresholds and
the critical exponents are summarized in Table~\ref{cridopuv}.

The ordinary bond percolation on the $(u,v)$ flowers was investigated in
Ref.~\cite{RozenfeldPRE2007}. Comparing the two studies, we find that the
DOP and the ordinary bond percolation belong to the same universality class
for $u\neq 1$. On the other hand, the critical behaviors are different when
$u=1$.
The ordinary bond percolation transition is of infinite order with
$\beta=\infty$, while the DOP transition is characterized with $\beta=1$.
Recently, the ordinary site percolation on $(1,2)$ flowers was studied
in Ref.~\cite{Hasegawa13}. It was found that the system is critical in the
region $0<p<1$ and the percolation transition takes place at $p=p_c=1$.
Thus, when $u=1$, the DOP transition is also distinct from the ordinary site
percolation transition.

It is interesting to note that the DOP and the ordinary percolation belong
to the same universality class when $u\neq 1$. The result is consistent
with the findings of Ref.~\cite{Gallos05} studying a percolation transition
of scale-free networks under a targeted attack. Each node $i$ with degree
$k_i$ fails with a probability $W(k_i)\propto k_i^\alpha$ with a
control parameter $\alpha$.
As the fraction of failed nodes varies, the giant
cluster of working nodes undergo a percolation transition. When $\alpha=0$,
it is equivalent to the ordinary site percolation.
For positive values of $\alpha$,
the larger degree a node has, the more vulnerable it is, and vice versa.
It was found in Ref.~\cite{Gallos05} that the percolation transitions with
negative $\alpha$ belong to the same universality class as the random
percolation transition~($\alpha=0$). The DOP corresponds to the case with
$\alpha=-\infty$. It is surprising that the extreme-case percolation
displays the same critical phenomena as the random percolation.

In Ref.~\cite{LeePRE2013}, we proposed that the epidemic threshold of the
SIS model on a network can be zero only when the threshold of the DOP on the
same network is zero.
The epidemic threshold was found to be zero at $u=1$ and nonzero
for $u\neq 1$ numerically~\cite{LeePRE2013}. The analytic results for the
DOP threshold for the $(u,v)$ flowers support the proposal. 
The DOP in general random scale-free networks is left for future work.

\begin{acknowledgments}
This work was supported by the Basic Science Research Program through the NRF
Grant No.~2013R1A2A2A05006776. This work was also supported by the NRF Grant
No.~2010-0015066.
\end{acknowledgments}

\end{document}